\documentclass[preprint]{elsarticle}
\usepackage{lineno,hyperref}
\usepackage{graphicx}
\usepackage{dcolumn}
\usepackage{bm}
\usepackage{longtable}
\journal{Physics Letters B}
\usepackage{graphicx}  
\graphicspath{ {figures/} } 
\usepackage{bm}     
\usepackage{amssymb}   
\usepackage{amsmath}   
\usepackage{hyperref}
\usepackage{ulem}
\usepackage{natbib}
\usepackage{footmisc}
\usepackage{xcolor}
\usepackage[utf8]{inputenc}
\usepackage{atbegshi}
\usepackage{comment}
\AtBeginDocument{\AtBeginShipoutNext{\AtBeginShipoutDiscard}}
\usepackage{lipsum}

\begin{document}
\setlength{\tabcolsep}{6pt}
\begin{frontmatter}
\title{Experimental Evidence of Large Collective Enhancement of Nuclear Level Density and its Significance in Radiative Neutron Capture}

\author[1,2]{T.~Santhosh}\ead{tsanthu@barc.gov.in}
\address[1]{Nuclear Physics Division, Bhabha Atomic Reserch Centre, Trombay, Mumbai- 400085, INDIA}

\address[2]{Homi Bhabha National Institute,Anushaktinagar,Mumbai- 400094, INDIA}

\author[1,2]{P.~C.~Rout}\ead{prout@barc.gov.in}
\author[1,2]{S.~Santra}
\author[1,2]{A.~Shrivastava}
\author[1]{G.~Mohanto}\
\author[1,2]{S.~K.~Pandit}
\author[1]{A.~Pal}
\author[1,2]{Ramandeep~Gandhi}
\author[1,2]{A.~Baishya}
\author[1,2]{Sangeeta~Dhuri}

\date{\today}

\begin{abstract}
The collective enhancement of nuclear level density and its fade out with excitation energy in deformed $^{171}$Yb nucleus has been inferred through an exclusive measurement of neutron spectra.The statistical model analysis of neutron spectra demonstrated a large collective enhancement factor of 40$\pm$3 for the first time, which corroborates with the recent microscopic model predictions but is an anomalous result compared with the measurements in the nearby deformed nuclei. The complete picture of the energy dependent collective enhancement has been obtained by combining with Oslo data below neutron binding energy. The significance of large collective enhancement in radiative neutron capture cross section of astrophysical interest is highlighted.

\end{abstract}
\end{frontmatter}

 

The atomic nucleus is a many body fermionic system manifesting nuclear properties due to both single particle and collective character of the nucleons. In the single particle picture, the spin-orbit interaction of the nucleons along with the mean-field potential explains the extra stability of the magic nuclei, while the coherent movement of a large number of nucleons exhibits collective excitation such as vibration and rotation. The rotation in nuclei also represents the collective mode associated with the spontaneous symmetry breaking due to deformation, related to Anderson, Goldstone and Nambu (AGN) bosons analogous excitations in superconductivity~\cite{BD02}. Many nuclear phenomena such as resonances, fission isomers, shape transition etc., are due to the coexistence of both single particle and collective degrees of excitation of nucleons in the nucleus~\cite{BM01}.

The collective modes in the atomic nuclei influence the Nuclear Level Density(NLD), a crucial physical quantity to provide the thermodynamic properties of the excited nucleus~\cite{TE}. The NLD is vital in various studies including low energy astrophysical reaction rates and nucleosynthesis in stars\cite{TR1,TR}. Accurate measurement of the collective contribution to the total NLD (collective enhancement) is crucial for statistical models like Hauser-Feshbach for the calculation of radiative capture rates significant for astrophysical interest. Although the NLDs have been studied extensively in the past, the effects of collective enhancement of the NLD for different mass regions are less understood both theoretically and experimentally.

The NLDs have been mainly obtained by inversion of Laplace transform of the grand partition function either by numerical methods or by saddle point approximation~\cite{TD}. The simplest form of single particle level density based on Fermi gas assumption was derived by Bethe~\cite{HB} and the energy(E) dependent level density is proportional to $ \frac{1}{E^2}$ {e$^{2\sqrt{aE}}$}, where \textbf{$a$} is the level density parameter with  $a = \pi^2g/6$, \textbf{$g$} is the single-particle level density evaluated at the Fermi surface.
The Fermi gas model is appropriate for explaining the level density at high energy while the low energy levels are dominated by correlation effects that give rise to pair correlations, structure effects such as shell effect, and collective modes of rotational and vibrational type superimposed on the single-particle motion. On the other hand, the shell model quantum Monte Carlo method(SMMC)~\cite{AO,YA} has been successfully applied to several nuclei to calculate NLDs, whereas Hartree-Fock BCS~\cite{PD},  Hartree-Fock-Bogolyubov plus combinatorial method \cite{SH} gives a global description of the NLDs. However, these microscopic models, if the particle number is not conserved, have to be renormalized with parameters extracted from the experimental analysis of the cumulative number of levels and s-wave neutron spacing at the neutron binding energy\cite{SG}.

The intrinsic level density for a spherical nuclei at a fixed angular momentum J, Excitation energy E is,
\[\rho(E_X,J)=\frac{(2J+1)}{\sqrt{8\pi}\sigma^{3}}e^{{\frac{-J(J+1)}{2\sigma^{2}}}}\rho_{int}(E_X),\]
Here the spin cut-off parameter $\sigma$ gives the width of the spin distribution and can be expressed as $\sigma=\sqrt{g\langle{m^2}\rangle T}$, Where g is single particle level density at fermi energy, $\langle{m^2}\rangle$ is the average of square of the spin distribution near fermi level, T is the nuclear temperature.

The level density for a deformed axially symmetric nucleus of specified spin J is obtained by summing over the intrinsic states with specified K (spin projection on symmetry axis), given as, 
\[\rho(E_{X},J)=\sum_{K=-J}^{J}\frac{1}{\sqrt{8\pi}\sigma_\perp}e^{{\frac{-K^2}{2\sigma^{2}}}}\rho_{int}(E_{X}-E_{rot}),\] 

 From the above expression, the level density for an axially symmetric deformed nuclei is enhanced by the factor of the order of $\sigma^{2}_\perp$ ($\sigma_\perp$ is the perpendicular spin cut off parameter)~\cite{BM01}. This is ranging from 35 to 65 for nuclei of A $\approx$ 150 to 250 at excitation energies around the neutron binding energy~\cite{Stokstad}. One of the recent works on NLD using finite-temperature relativistic Hartree-Bogoliubov model \cite{jz}, showed that the enhancement in the mass region $A$ = 160-170 is $\approx$ 40. In another work~\cite{smg}, using a microscopic level density model, a similar magnitude of collective enhancement has been reported. It is shown that, in a fully deformed nucleus, the vibrational enhancement factor is small($\sim$2) compared to rotational enhancement factor($\sim$10-100)~\cite{BBM}. 

So far, several experiments have been reported to address the collective enhancement of the NLD and its fade-out. However, all the recent experimental evidences on the collective enhancement factor are well below the predicted value. Jhungans{\it { et al.}}~\cite{Ajung} first attempted to study the fade-out of collective enhancement in a projectile fragmentation experiment and the inclusion of collective enhancement of the NLD was required to explain the experimental data. Komarov{\it { et al.}}~\cite{Skom} searched for fade out of collective enhancement in the energy region 30-60 MeV as predicted by SU3 shell model \cite{HJ} but found no such evidence. This null result could be due to the fade out of collective enhancement much earlier than predicted as observed in Refs. \cite{AO,YA}. Recently, the statistical model analysis of neutron evaporation  spectra~\cite{KB} from alpha induced reaction populating $^{172}$Lu and $^{184}$Re at excitation energy 22-56 MeV inferred the fade out of collective enhancement. In a coincidence experiment~\cite{GM}, the observed enhancement factor in the mass A$\sim$190 region was found to be $\sim$8 and its fade-out was at $\sim$14~MeV. A similar collective enhancement factor of 10 was obtained from the statistical model analysis of neutron, proton and gamma-ray spectra measured from the deformed $^{169}$Tm and $^{185}$Re populated at excitation energy around 26~MeV~\cite{DP} and the enhancement was disappeared around 14 MeV. These experimental contradictions made the understanding of the contribution of collective degrees of freedom to the intrinsic level density a puzzle of nuclear physics. There is little consensus as to whether the collective enhancement for deformed nuclei is as high as $\sigma^{2}_\perp$. 

\begin{figure}[t]
    \centering
    \includegraphics[width=0.5\textwidth]{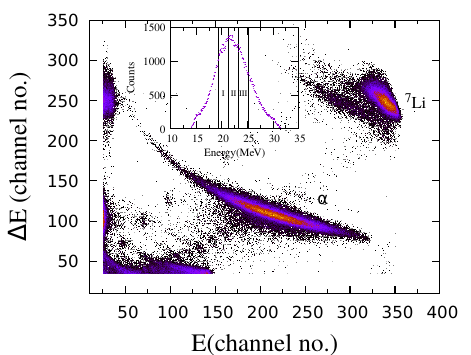}
    \caption{A typical particle energy in $\Delta$E as a function of energy E in $\Delta$E-E Si-Strip telescope used to identify charged particles using Bethe-Bloch formula. The inset shows the projected alpha energy spectrum and also indicated the three energy bins. }
    
    \label{fig:01}
\end{figure}

\begin{figure}[h]
    \centering
      \includegraphics[width=0.5\textwidth]{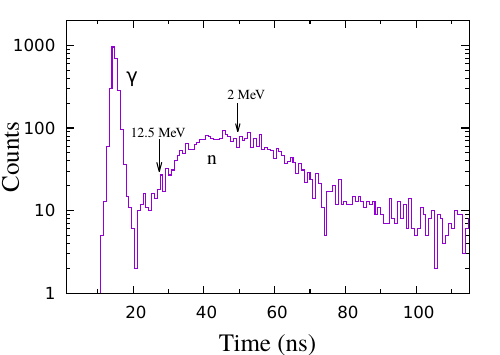}
    \caption{Time of flight spectrum in $^7$Li + $^{169}$Tm reaction for the central energy bin
of alpha particles. The arrows indicate the positions for two representative neutron energies.}
        \label{fig:02}
\end{figure}

In this letter, we present an exclusive measurement of neutron spectra from $^{172}$Yb following the triton transfer in the $^{7}$Li+ $^{169}$Tm reaction, in coincidence with ejected alpha particles using the similar technique as in ref.~\cite{pcr1} with improved experimental setup. The nucleus $^{172}$Yb (formed at 25.5-27.5~MeV corresponding to selected alpha energies) decays predominantly through neutron emission populating the residual nucleus, where the fade-out of collective enhancement is expected to show significant changes in neutron spectra. Study of level density in the mass region A $\sim$~170 is also important for s-process branch point in low-mass asymptotic giant branch (AGB) stars~\cite{NTOF}.

The experiment was performed at the BARC-TIFR Pelletron Linac Facility (Mumbai) using 40~MeV pulsed $^{7}$Li beam of width $\sim$1.5~ns (FWHM) and period $\sim$107~ns. The pulsed beam was bombarded on a self-supported 2.72 mg/cm$^2$-thick $^{169}$Tm target with an average beam current of 12~enA. The evaporated neutrons from the excited $^{172}$Yb nucleus were detected in coincidence with the outgoing alpha particles.
The charged particles were detected using four $\Delta$E-E telescopes of double-sided silicon strip detectors(DSSD) of 5~cm$\times$5~cm dimension. These detectors were kept at 10 cm away from the target center at mean angles $\pm$60$^\circ$ and $\pm$140$^\circ$ with respect to the beam direction. The thickness of $\Delta$E and E detectors was 50$\mu$m and 1500$\mu$m, respectively. Each detector consists of 16 strips back and front and covers an angular range of 25$^\circ$. Signal read-outs were taken from each strip. In order to detect evaporated neutrons, an array of 15 liquid scintillation(LS, EJ301) detectors arranged in circular geometry~\cite{pcr2} were used. Each LS detector was cylindrical with a diameter 12.5~cm and a thickness 5~cm. Three rows of LS detectors were stacked to form an array and each detector was placed at $\sim$72~cm from the target center. The array covered an angular range from  58$^\circ$ to 143$^\circ$. There was a 16$^\circ$ separation between each detector.
Each LS detector was coupled to a fast photo-multiplier tube(PMT) of diameter 12.5~cm for signal readouts. The unambiguous detection of neutrons amidst the gamma background is possible due to both time of flight (TOF) and pulse shape discrimination (PSD) methods employed for the LS detectors. The TOF, PSD, pulse height energy of each LS detector and energy signal from each strip detector has been recorded in list mode using a VME based data acquisition system. The TOF was calibrated using a precision time calibrator, and the pulse height calibration was done by identifying the Compton edge from $^{137}$Cs, $^{22}$Na and Am-Be sources. The energy calibration of each strip detector from 4.8~MeV to 8.5~MeV was performed using $^{229}$Th and $^{241}$Am sources and assuming linear calibration thereafter up to 30~MeV. The shadow pyramid built using several iron plates of thickness 30cm has been used to assess the scattered contribution to direct neutron emission from the reaction. The beam dump was shielded with several layers of borated paraffin and lead blocks to reduce the neutrons and gamma rays background.
\begin{figure}[t]
\begin{center}

\includegraphics[width=0.47\textwidth]{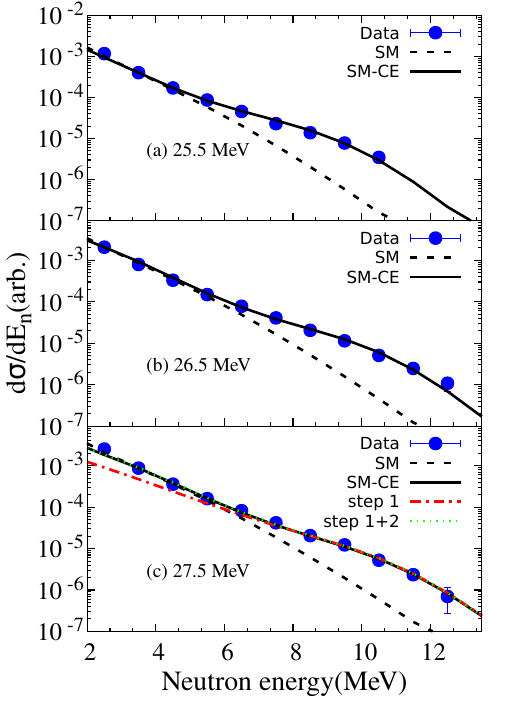}

\end{center}
\caption{Comparison of neutron spectra with Statistical Model calculation using the level density parameter A/8.5 MeV$^{-1}$. Solid line show calculation with collective enhancement (SM-CE) and dashed line is without collective enhancement (SM). Three excitation energies (a) 25.5~MeV, (b) 26.5~MeV and (c) 27.5~MeV, respectively, for three alpha energy gates(III, II and I) as shown in inset of the Fig.~\ref{fig:01}. (d) Shows the contributions from step 1 and step 1+2 for the energy 27.5 MeV.}.
\label{fig:ns}
\end{figure}

\begin{figure}[t]
\begin{center}
\includegraphics[width=0.5\textwidth]{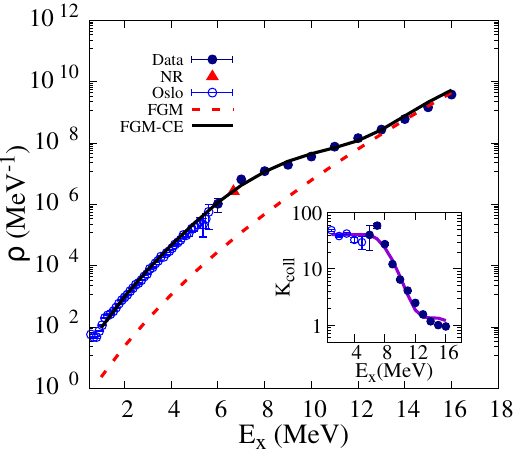}

\end{center}
\caption{The NLD obtained as a function of excitation energy using Oslo data(open circles)~\cite{AS} and present experiment(filled circles), normalized to the level density at the neutron resonance point(NR)(triangle). The dashed red line shows intrinsic level density from fermi gas model(FGM) and black solid line shows fermi gas level density with collective enhancement(FGM-CE). The inset shows collective enhancement with excitation energy obtained from the combined data sets.}
\label{fig:rho}
\end{figure}

Alpha particles were clearly separated from the elastic $^7$Li as shown in Fig.~\ref{fig:01} by utilizing Bethe-Bloch energy loss technique in $\Delta$E-E telescope. The inset of Fig.~\ref{fig:01} is the projected alpha energy spectrum depicted with three energy bins. A typical TOF spectrum is shown in Fig.~\ref{fig:02}. The prompt timing signal of gamma rays was used as a reference to find the absolute neutron time of flights. The random backgrounds were subtracted from the TOF spectra while deriving the neutron energy spectra. It is worth mentioning that scattered neutrons from the surroundings have been assessed by placing several layers of Iron plates shadow pyramid between the detector and target. The scattered neutron contribution was found to be less than 1\% in the energy region of interest. The efficiency of the LS detector as a function of incident neutron energy and threshold energy was calculated using a Monte Carlo simulation~\cite{pcr2}, which was validated using measured neutrons from $^7$Li (p,n) reaction. The efficiency corrected energy spectra of neutrons was derived from the TOF spectrum with a suitable gate from the PSD spectrum. For neutrons of 12 MeV energy, an energy resolution(FWHM) of 1 MeV was achieved. The neutron spectra derived for all three alpha energy gates as given in the inset of Fig.~\ref{fig:01} corresponding to the excitation energy of $^{172}$Yb of 25.5, 26.5 and 27.5~MeV are shown in the Fig.~\ref{fig:ns}. The forward and backward spectra in centre of mass frame are found to be symmetric with respect to 90$^\circ$ indicating that the neutron emission is from a statistically equilibrated system.

The statistical model~(SM) analysis of the neutron spectra was done using the code CASCADE \cite{puhl} with the E$_X$ and J dependent NLD,
\[\rho(E_X,J)=\frac{(2J+1)\sqrt{a}}{12U^2}\left(\frac{\hbar^2}{2\Im}\right)^{3/2}e^{2\sqrt{aU}},\]
where $U = E_X - E_{\rm{rot}} - \Delta_P$, $\Delta_P$ is the 
pairing energy and $E_{\rm{rot}} = \left(\frac{\hbar^2}{2\Im}\right) J(J+1), \Im$ being the moment of inertia. The excitation energy dependence of the NLD parameter $a$, which includes the shell effect and its damping, has been parameterized~\cite{pcr1,avi} as \[a=\tilde{a}[1-\frac{\Delta_S}{U}(1 - e^{-\gamma U})].\] Here $\tilde{a}$ is the asymptotic value of the NLD parameter in the liquid drop region, $\Delta_S$  is the shell correction energy, which is the difference between the experimental binding energy and that calculated from the LDM, and $\gamma$ is the damping parameter.

The single particle and the collective contributions can be decoupled and the Phenomenological expression for the total NLD can be written as intrinsic level density multiplied by a collective factor,  \[ \rho_{tot} = \rho_{int} K_{coll},\] Where $K_{coll}$ includes both rotational and vibrational enhancement contribution.

In our present study, the CASCADE model was modified to include the collective enhancement factor $K_{coll}$ as fermi function given as, 
\[ K_{coll} = 1+A_{en}.{\frac{1}{1+e^{\frac{(E-E_{cr})}{D_{cr}}}}}
~,\]
where A$_{en}$ is the maximum collective enhancement factor, E$_{cr}$ is the critical energy where the enhancement drops to half, D$_{cr}$ is the width of the transition region.

As the excitation energy increases over a certain range, the deformed core should not be able to contribute collective degrees of freedom to the NLD due to large mixing of incoherent excitations and symmetry breaking will be compromised. Consequently, collective enhancement should gradually decrease with increasing excitation energy, and above this the total NLD has the contribution only from intrinsic states. This phase transition of enhancement from the collective modes should be reflected in the number of neutrons emitted populating the residual nuclei, giving variation in the slope of the evaporation spectra. 

The statistical model analysis of measured neutron spectra with the level density parameter A/8.5 MeV$^{-1}$ with and without collective enhancement have been carried out.Figure~\ref{fig:ns}(a), (b), and (c) show the statistical model(CASCADE) calculation with collective enhancement~(SM-CE) and without collective enhancement~(SM) for the Yb corresponding to three excitation energies 25.5, 26.5 and 27.5 MeV. The contribution of evaporation from multiple steps was estimated for the highest energy of 27.5 MeV. As shown in Fig. 3(c), Above 6 MeV neutron energy where collective enhancement is observed, the contribution is mainly from first step evaporation. The energy covered for the residual nuclei is 6-18 MeV where the fade-out of collective contribution is observed. It is inferred from the SM analysis of the measured neutron spectrum that the collective enhancement factor for $^{171}$Yb is 40. The sensitivity of level density parameter was also studied by varying the inverse level density parameter with k= 8.5$\pm$0.2 MeV. The collective enhancement and fade-out energy were found to be 40$\pm$3 and 14$\pm$1 MeV respectively by analysis all three excitation energies. We have used the shell correction energy $\Delta{s}\sim$~0.34 MeV and $\gamma$ =0.005 for the present calculation used in the CASCADE code, taken from the Ref. \cite{wdm,pcr1}. Interestingly, the observed enhancement factor is considerably larger than the several other recent measurements in near mass region~\cite{KB,GM,DP} but in agreement with recent microscopic level density calculations\cite{jz,smg}.

In order to get the complete picture of the collective enhancement and its fade out, we have combined the level density of $^{171}$Yb from Oslo data~\cite{AS} with the present measurement as shown in Figure~\ref{fig:rho}. The Oslo method level density ranged up to 5.6 MeV($\sim$ B$_{n}$-1, B$_{n}$ is neutron binding energy), while our measurement covered above 6 MeV. The experimental level density from the present measurement was obtained for the excitaion energy 26.5 MeV by using the following scaling technique as described for proton evaporation spectrum in Ref.~\cite{DRC}.
\[\rho_{exp}(E_X)=\frac{(d\sigma/dE_{n})_{exp}}{(d\sigma/dE_{n})_{fit}}\rho_{fit}(E_X),\]

Then the derived level density was normalized at neutron resonance energy taken from Ref.~\cite{NR} and combined with Oslo NLD of $^{171}$Yb to obtain level density from $\sim$~1~MeV to 16~MeV~(Figure~\ref{fig:rho}). The inset of Figure~\ref{fig:rho} shows the collective enhancement as a function of excitation energy obtained from the ratio between measured NLD and the Fermi gas level density. It is also evident from the figure that the enhancement factor is $\sim$40 and fade-out energy is $\sim$14 MeV. It can be worth mentioning that both Oslo data and neutron data give the same enhancement factor with respect to intrinsic level density given by the Fermi gas model. This observation has important significance in astrophysical neutron capture rates.

\begin{figure}[t]
\begin{center}
\includegraphics[scale=1.1]{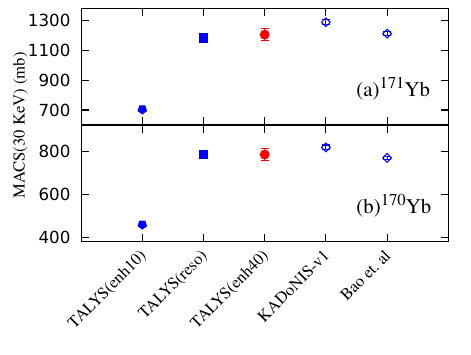}

\end{center}
\caption{(a) Calculated MACS at 30 KeV for $^{171}$Yb using the Talys-1.96 and its comparison with KADoNIS-v1 compilation, well-established Bao \textit{et al.} estimation, (b) Same as (a) for $^{170}$Yb. Here, TALYS(enh10) and TALYS(enh40) were using the present level density prescription with enhancement factor 40 and 10, respectively. TALYS(reso) was the calculated MACS using the measured level density at neutron resonance energy.}
\label{fig:macs}
\end{figure}

The neutron-capture process is responsible for the formation of the heavy nuclei between iron and the actinides~\cite{FK}. In order to find the implication of large collective enhancement, we have calculated Maxwellian average neutron capture cross section (MACS) at 30~KeV for $^{171}$Yb and $^{170}$Yb. This is achieved by incorporating our collective enhancement form in TALYS-1.96 reaction code~\cite{talys21}. The level density of Fermi gas model with the measured collective enhancement used while calculating MACS. The calculated MACS values were then compared with KADoNIS-v1~\cite{kad}, estimation of Bao \textit{et al.}~\cite{Bao} for both $^{171}$Yb and $^{170}$Yb as shown in Fig.~\ref{fig:macs} (a) and (b). To see the effect of collective enhancement on MACS, various TALYS calculations with different collective enhancement factor has been performed. Fig.~\ref{fig:macs} shows the MACS calculated using TALYS with enhancement factors 40 and 10. The predicted MACS using the present level density prescription including the measured enhancement factor(TALYS(enh40)) agrees with the experimental MACS values while the collective enhancement factor 10 (TALYS(enh10)) could not reproduce well. Therefore, in a statistical model, it is necessary to include the proper NLD prescription with collective enhancement which significantly improves the predicted capture cross sections relevant for astrophysical s-process.

In conclusion, we have inferred the collective enhancement of NLD through the statistical analysis of the measured neutron evaporation spectrum in the mass A~$\sim$170 region where this effect is significant. The experimental results show that the collective enhancement factor of 40~$\pm$~3 in $^{171}$Yb, is the largest collective enhancement factor reported in any system to date and the fade out energy is 14$\pm$1~MeV. The complete form of energy dependent collective enhancement was experimentally deduced for the first time by combining the measured NLD with the Oslo data. It is also found that suitable level density prescription with appropriate collective enhancement has significantly improves the prediction of stellar neutron capture cross sections. The results on the collective enhancement are important for research in the synthesis of super-heavy elements involving deformed targets and also relevant for s-process branch points in AGB stars. 

We thank V.M. Datar and D.R. Chakrabarty for their valuable comments and suggestions, late R. Kujur for his help during the experiment. The authors, TS and SD are sincerely grateful to DST for financial support under the DST-INSPIRE Fellowship scheme.


\begin{thebibliography}{50}

\bibitem{BD02} D. Brink, Nuclear Superfluidity - Pairing in Finite Systems, CUP, 2005.
\bibitem{BM01} A. Bohr and B. R. Mottelson, Nuclear Structure, 1st ed. (Benjamin, 1969).
\bibitem{TE}T. Ericson, Adv. Phys. 9, 425 (1960).
\bibitem{TR1}T. Rauscher, F. K. Thielemann, and K. L. Kratz, Phys. Rev. C
56, 1613 (1997).
\bibitem{TR}T. Rauscher and F. K. Thielemann, At. Data Nucl. Data
Tables 75, 1 (2000).
\bibitem{TD}T. Dossing and A. S. Jensen, Nucl. Phys. A222, 493 (1974).
\bibitem{HB}H.A.Bethe,Phys. Rev. 50, 332 (1936).
\bibitem{AO}  A. Ozen, Y. Alhassid, H. Nakada, Phys. Rev. Lett. 110,042502 (2013).
\bibitem{YA} Y. Alhassid, G. F. Bertsch, C. N. Gilbreth, and H. Nakada, Phys.Rev.C93, 044320 (2016).
\bibitem{PD}P. Demetriou and S. Goriely, Nucl. Phys. A695, 95 (2001).
\bibitem{SH}S. Hilaire and S. Goriely, Nucl. Phys. A779, 63 (2006).
\bibitem{SG}S. Goriely, S. Hilaire, and A. J. Koning, Phys. Rev. C 78, 064307 (2008)
\bibitem{Stokstad} Stokstad, R.,  The use of the statistical Model in Heavy-ion Reaction studies, LBNL Report No: LBL-12636, (1981).
\bibitem{jz}Jie Zhao, Tamara Niksic, and Dario Vretenar,  Phys. Rev. C 102, 054606(2020).
\bibitem{smg} S. M. Grimes, T. N. Massey, and A. V. Voinov, Phys. Rev. C 99, 064331 (2019).
\bibitem{BBM} S. Bjørnholm, A. Bohr, B. Mottelson, Physics and Chemistry of Fission, Proceedings of a Conference at Rochester,
Vol. 1 (IAEA, Vienna, 1974) p. 367.
\bibitem{Ajung} A. Junghans, M. de Jong, H.-G. Clerc, A. Ignatyuk, G. Kudyaev, and K.-H. Schmidt, Nucl. Phys. A 629, 635 (1998).
\bibitem{Skom} S. Komarov, R. J. Charity, C. J. Chiara, W. Reviol, D. G. Sarantites, L. G. Sobotka, A. L. Caraley, M. P. Carpenter, and D. Seweryniak, Phys. Rev. C 75, 064611 (2007).
\bibitem{HJ}G.Hansen and A.S.Jensen, Nuclear Physics A406(1983)236
\bibitem{DA}T. Døssing, S. Åberg, Eur. Phys. J. A (2019) 55: 249
\bibitem{KB} K. Banerjee, P. Roy, D. Pandit, J. Sadhukhan, S. Bhattacharya, C. Bhattacharya, G. Mukherjee, T. Ghosh, S. Kundu, A. Sen, T. Rana, S. Manna, R. Pandey, T. Roy, A. Dhal, M. Asgar, and S. Mukhopadhyay, Phys. Lett. B 772, 105 (2017).
\bibitem{GM} G. Mohanto, A. Parihari, P. C. Rout, S. De, E. T. Mirgule, B. Srinivasan, K. Mahata, S. P. Behera, M. Kushwaha, D. Sarkar, B. K. Nayak, A. Saxena, A. K. Rhine Kumar, A. Gandhi, Sangeeta, Nabendu K. Deb, and P. Arumugam, Phys. Rev. C 100 (2019) 011602(R).
\bibitem{DP} Deepak Pandit, Balaram Dey, Srijit Bhattacharya, T.K. Rana, Debasish Mondal, S. Mukhopadhyay, Surajit Pal, A. De, Pratap Roy, K. Banerjee, Samir Kundu, A.K. Sikdar, C. Bhattacharya, S.R. Banerjee, Phys. Lett. B 816, 136173 (2021).

\bibitem{pcr1}P. C. Rout, D. R. Chakrabarty, V. M. Datar, S. Kumar, E. T. Mirgule, A. Mitra, V. Nanal, S. P. Behera, and V. Singh, Phys. Rev. Lett. 110, 062501 (2013).
\bibitem{NTOF}C. Guerrero et. al.,nTOF collaboration, Phys. Rev. Lett. 125, 142701 (2020).
\bibitem{pcr2}P.C. Rout, A. Gandhi, T. Basak, R.G. Thomas, C. Ghosh, A. Mitra, G. Mishra, S.P. Behera, R. Kujur, E.T. Mirgule, B.K. Nayak, A. Saxena, S. Kumar, and V.M. Datar, J. Inst. 13, P01027 (2018).
\bibitem{puhl}F. Puhlhofer, Nucl. Phys. A280, 267 (1977)
\bibitem{avi}A. V. Ignatyuk, G. N. Smirenkin, and A. S. Tishin, Sov. J. Nucl. Phys. 21, 255 (1975).
\bibitem{AS}A. Schiller, A. Bjerve, M. Guttormsen, M. Hjorth-Jensen, F. Ingebretsen, E. Melby, S. Messelt, J. Rekstad, S. Siem, and S. W. Ødegård, Phys. Rev. C 63, 021306(R) (2001).
\bibitem{wdm} W.D. Myers, W.J. Swiatecki, Lawrence Berkeley Laboratory Report No. LBL-36803, 1994.
\bibitem{DRC} D. R. Chakrabarty, V. M. Datar, Suresh Kumar, E. T. Mirgule, H. H. Oza, and U. K. Pal,Phys. Rev. C 51, 2942 (1995).
\bibitem{NR}Handbook for Calculations of Nuclear Reactions Data, IAEA, Vienna, Report No. IAEA-TECDOC-1024, 1998.
\bibitem{FK} F. Käppeler, R. Gallino, S. Bisterzo, and W. Aoki, Rev. Mod. Phys. 83, 157 (2011).
\bibitem{talys21} A. Koning, S. Hilaire, S. Goriely  TALYS-1.9 A Nuclear Reaction Program. User Manual; Nuclear Research and Consultancy Group (NRG): Petten, The Netherlands, 2015.
\bibitem{kad} KADoNiS-The Karlsruhe Astrophysical Database of Nucleosynthesis in Stars, online at  https://exp-astro.de/kadonis1.0
\bibitem{Bao} Z. Y. Bao, H. Beer, F. Käppeler, F. Voss, K. Wisshak, and
T. Rauscher, At. Data Nucl. Data Tables 76, 70 (2000).

\end{thebibliography}
\end{document}